\documentclass[pra,showpacs]{revtex4}
\usepackage{amsmath}
\usepackage{epsfig}
\usepackage{array}
\usepackage{natbib}

\begin{document}

\title{Manifestations of nuclear anapole moments in solid state NMR}

\author{T.N. Mukhamedjanov, O.P. Sushkov, J.M. Cadogan}

\affiliation{School of Physics, University of New South Wales,\\
 Sydney 2052, Australia}

\begin{abstract}
We suggest to use insulating garnets doped by rare earth ions for measurements of
nuclear anapole moments. A parity violating shift of the NMR frequency arises due to the combined effect of the lattice crystal field and the anapole moment of the rare-earth nucleus.

We show that there are two different observable effects related to frequency: 
1) A shift of the NMR frequency in
an external electric field applied to the solid. The value of the shift is
about $\Delta \nu_1 \sim 10^{-5}\ \textrm{Hz}$ with $E=10\ \textrm{kV/cm}$;
2) A splitting of the NMR line into two lines. The second effect is independent of
the external electric field. The value of the splitting is about
$\Delta \nu_2 \sim 0.5\ \textrm{Hz}$ and it depends on the orientation of the crystal with
respect to magnetic field. Both estimates are presented for a magnetic field of
about 10 tesla.

We also discuss a radiofrequency electric field  and a static macroscopic
magnetization caused by the nuclear anapole moment.
\end{abstract}

\pacs{11.30.Er, 21.10.Ky, 71.15.Dx}

\maketitle

\section{Introduction}
The anapole moment is a characteristic of a system which is related to the toroidal
magnetic field confined within the system. It was pointed out some time ago by Zeldovich \cite{Zeld}
that the anapole moment is related to parity violation inside the system.
Interest in the nuclear anapole moment is mostly due to the fact that it gives dominating contribution
to effects of atomic parity nonconservation (PNC) which depend on nuclear spin \cite{FK}.
There are two mechanisms that contribute to these effects. The first is due to exchange of 
a $Z$-boson between electron and nucleus. The second mechanism is due to the usual magnetic interaction of an electron with the nuclear anapole moment.
The contribution of the first mechanism is proportional to $1-4s^2$.
Since sine squared of the Weinberg angle is $s^2\approx0.23$ \cite{PD}, the first mechanism is 
strongly suppressed and the second mechanism dominates.
The anapole moment of $^{133}$Cs has been measured in an optical PNC experiment with atomic Cs \cite{Cs}.
This is the only observation of a nuclear anapole moment.
There have been several different suggestions for measurements of nuclear anapole moments. 
Measurements in optical transitions in atoms or in diatomic molecules remains an option, for a review
see \cite{Khripl}.
Another possibility is related to radiofrequency (RF) transitions in atoms or diatomic molecules 
\cite{NK,LS,L1,SF}.
Possibilities to detect nuclear anapole moments using collective quantum effects
in superconductors \cite{VK}, as well as PNC electric current in ferromagnets \cite{Labz},
have been also discussed in the literature. 
A very interesting idea to use Cs atoms trapped in solid $^{4}$He has been recently suggested in
Ref.~\cite{Bouchiat}.

Our interest in the problem of the nuclear anapole moment in solids was stimulated by the recent suggestion for
searches of electron electric dipole moment in rare earth garnets \cite{Lam}.
Garnets are very good insulators which can be doped by rare earth ions. They are widely used for
lasers and their optical and crystal properties are very well understood. To be specific we consider two 
cases: the first is yttrium aluminium garnet (YAG) doped by Tm \cite{Tm}. Thulium 3+ ions substitute 
for yttrium 3+ ions. The second case is yttrium gallium garnet doped by Pr \cite{Pr}.
Once more, praseodymium 3+ ions substitute for yttrium 3+ ions.
The dopant ions have an uncompensated electron spin $\boldsymbol{J}$
and a nuclear spin $\boldsymbol{I}$. For Tm$^{3+}$ $J=6$ and $I=1/2$ ($^{169}$Tm, 100\% abundance). For Pr$^{3+}$
$J=4$ and $I=5/2$ ($^{141}$Pr, 100\% abundance).

The simplest $P$-odd and $T$-even correlation ($P$ is space inversion and $T$ is time reflection) which arises due
to the nuclear anapole moment is
\begin{equation}
H_{\textrm{\it eff}}^{(1)} \propto [\boldsymbol{I} \times \boldsymbol{J}] \cdot \boldsymbol{E} ,
\label{HdynI}
\end{equation} 
where $\boldsymbol{E}$ is the external electric field.
It is convenient to use the magnitude of the effect expected in the electron electric dipole moment (EDM)
experiment \cite{Lam} as a reference point. For this reference point we use a value
of the electron EDM equal to the present experimental limit \cite{Com},
$d_e = 1.6\times 10^{-27} e \ \textrm{cm}$.
According to our calculations, the value of the effective interaction (\ref{HdynI}) is such that at 
the maximum possible value of the cross product $[\boldsymbol{I} \times \boldsymbol{J}]$
it induces an electric field four orders of magnitude higher than the electric field  
expected in the  EDM experiment \cite{Lam,MDS}.
For example, in Pr$_{3}$Ga$_{5}$O$_{12}$ the field is $E\sim 1.5 \times 10^{-6}\ \textrm{V/cm}$.
The problem is how to provide the
maximum cross product $[\boldsymbol{I} \times \boldsymbol{J}]$. Value of $\langle \boldsymbol{J}\rangle $
is proportional to the external magnetic field $\boldsymbol{B}$. A magnetic field of about $5$--$10\ \textrm{T}$ is
sufficient to induce the maximum magnetization. Nuclear spins can be polarized in the perpendicular direction
by an RF pulse, but then they will precess around the magnetic field with a frequency of about $1\ \textrm{GHz}$. It is not
clear if the anapole-induced voltage of this frequency can be detected.
An alternative possibility is to detect the static variation of the perpendicular magnetization induced by 
the external electric field, $\delta \boldsymbol{I} \propto  [\boldsymbol{B} \times \boldsymbol{E}]$.
The magnetization effect for Pr$_{3}$Ga$_{5}$O$_{12}$ is several times larger
than that expected for the EDM experiment \cite{Lam}.
This probably makes the magnetization effect rather promising.
In the present work we concentrate on the other possibility which is based on the crystal field of
the lattice. Because of the crystal field, the electron polarization of the rare earth ion has 
a component orthogonal to the magnetic field 
$\langle \boldsymbol{J}\rangle \propto \boldsymbol{B}+(\boldsymbol{B}\cdot \boldsymbol{n})\boldsymbol{n}$,
where $\boldsymbol{n}$ is some vector related to the lattice. The equilibrium orientation of the nuclear spin 
is determined by the direct action of the magnetic field together with the hyperfine interaction proportional
to $\langle \boldsymbol{J}\rangle$. Because of the $(\boldsymbol{B}\cdot \boldsymbol{n})\boldsymbol{n}$
term in $\langle \boldsymbol{J}\rangle$, the nuclear and the electron spins are not collinear, and the cross
product $[\boldsymbol{I} \times \boldsymbol{J}]$ is nonzero 
$[\boldsymbol{I} \times \boldsymbol{J}]\propto (\boldsymbol{B}\cdot \boldsymbol{n})
[\boldsymbol{B}\times \boldsymbol{n}]$. We found that NMR
frequency shift due to the correlation (\ref{HdynI}) is about 
\begin{equation}
\label{d1}
\Delta \nu_1 \sim 10^{-5}\ \textrm{Hz}
\end{equation}
at $E=10\ \textrm{kV/cm}$ and $B=10\ \textrm{T}$. In essence, we are talking about 
the correlation
$ (\boldsymbol{B}\cdot \boldsymbol{n})[\boldsymbol{B}\times \boldsymbol{n}]\cdot\boldsymbol{E}$ 
considered previously in the work of  Bouchiat and Bouchiat \cite{Bouchiat} for Cs trapped in solid $^{4}$He .

Another effect considered in the present work is the splitting of the NMR line into two lines due to
the nuclear anapole moment. This effect is related to the lattice structure and is independent of the 
external electric field.

The garnet lattice has a center of inversion. However, the environment of
each rare earth ion is asymmetric with respect to inversion. One can imagine that there is a microscopic
helix around each ion. Since the lattice is centrosymmetric, each unit cell has equal numbers of rare earth
ions surrounded by right and left helices (there are 24 rare earth sites within the cell). 
The microscopic helix is characterized by a third rank tensor $T_{klm}$ (lattice octupole). Together with the
nuclear anapole
interaction this gives a correlation similar to (\ref{HdynI}), but the effective ``electric field''
is generated now by the helix ${E}_k\propto T_{klm}J_lJ_m$. So the effective interaction is
\begin{equation}
H_{\textrm{\it eff}}^{(2)} \propto \epsilon_{ijk}I_iJ_jT_{klm}J_lJ_m .
\label{HstI}
\end{equation} 
The effective interaction (\ref{HstI}) produces a shift of the NMR line.
The value of the shift is about $0.5\ \textrm{Hz}$ at $B=10\ \textrm{T}$, and the sign of the shift is opposite for sites of different
``helicity'', so in the end it gives a splitting of the NMR line
\begin{equation}
\label{d2}
\Delta \nu_2 \sim 0.5\ \textrm{Hz}. 
\end{equation}
The value of the splitting depends on the orientation of the crystal with respect to the magnetic field. This is 
the ``handle'' which allows one to vary the effect.
Generically this effect is similar to the PNC energy shift in helical molecules \cite{molec}.

One can easily relate the values of the frequency shift in the external field (\ref{d1}) and of the line
splitting (\ref{d2}). The splitting  is due to internal atomic electric field which is about $10^9\ \textrm{V/cm}$.
Therefore, naturally, it is about 5 orders of magnitude larger than the shift (\ref{d1}) in field 
$10\ \textrm{kV/cm}$.

For the present calculations we use the jelly model suggested in Ref.~\cite{MDS}. Values of the nuclear anapole
moments of $^{169}$Tm and $^{141}$Pr which we use in the present paper have been calculated separately
\cite{MS}.
The structure of the present paper is as follows. In Section II the crystal structure of the compounds 
under consideration is discussed. The effective potential method used in our electronic structure 
calculations is explained in Section III.
The most important parts of the work which contain the calculations of the effective Hamiltonians 
(\ref{HdynI}) and (\ref{HstI}) are presented in Sections IV and V.
The crystal field and the angle between the nuclear and the electron spin is considered in 
Section VI. In section VII we calculate values of observable effects and  section VIII
presents our conclusions. Some technical details concerning the numerical solution of the equations for
electron wave functions are presented in Appendix.

\section{Crystal structure of Y($\mathbf{Pr}$)GG and Y($\mathbf{Tm}$)AG}

The compounds under consideration are ionic crystals consisting of Y$^{3+}$, O$^{2-}$, Ga$^{3+}$ ions for 
YGG and Al$^{3+}$ instead of Ga for YAG, plus Pr$^{3+}$ or Tm$^{3+}$ rare-earth doping ions. The chemical 
formula of YGG is Y$_{3}$Ga$_{5}$O$_{12}$ and the formula of YAG is Y$_{3}$Al$_{5}$O$_{12}$.
Yttrium gallium garnet and yttrium aluminium garnet belong to the $Ia\overline{3}d$ space group and contain 8 formula 
units per unit cell. Detailed structural data for these compounds are presented in Table \ref{strdata} 
\cite{YGGstr, YAGstr}.

\begin{figure}[htb]
\centering
\epsfig{figure=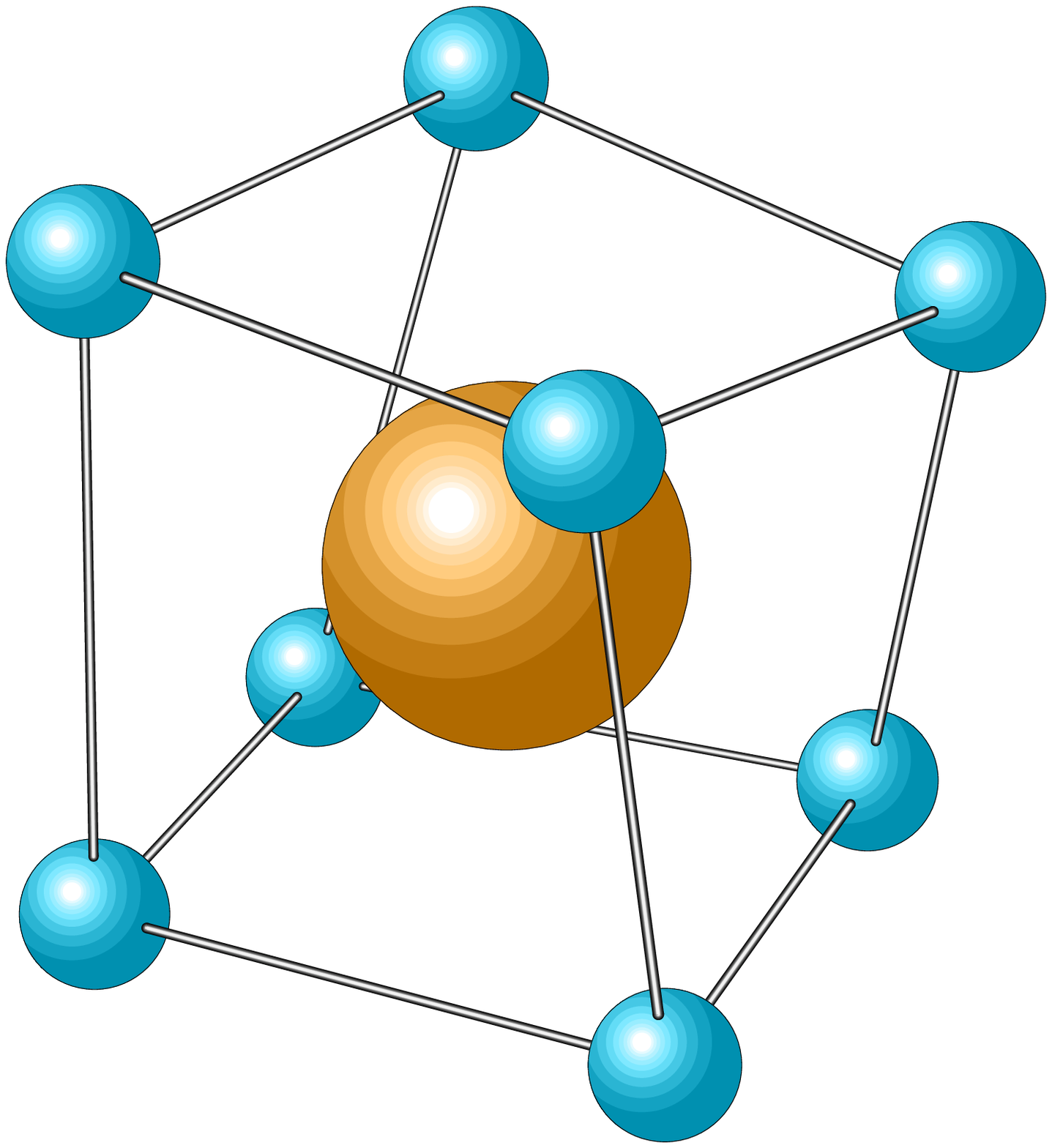,height=4cm}
\hspace{1cm}
\epsfig{figure=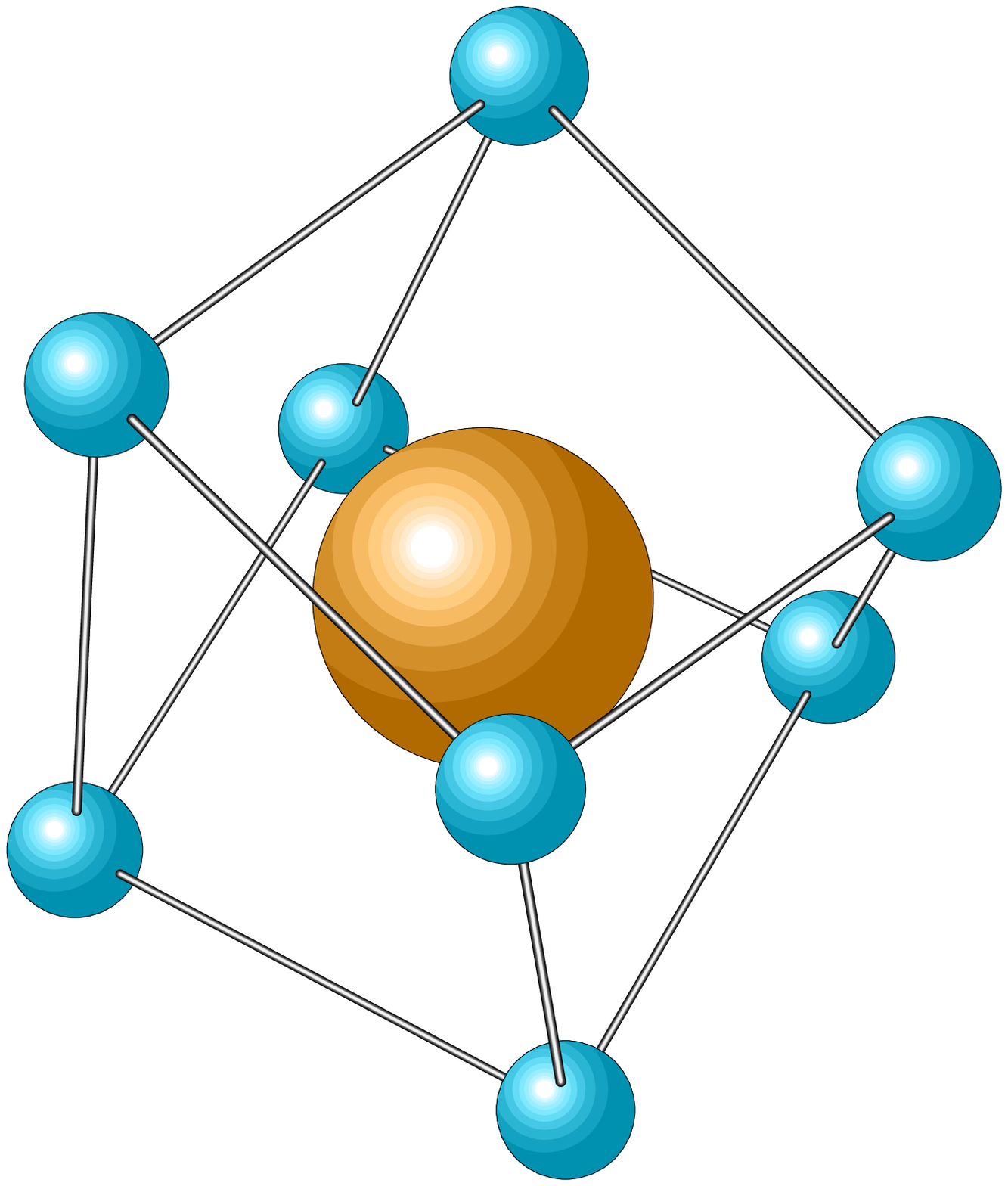,height=4.2cm}
\caption{
Dodecahedron configuration of O$^{2-}$ ions around the RE$^{3+}$ impurity ion in the
garnet structure. Two different viewing angles are shown.}
\label{garn}
\end{figure}

\begin{table}[h!]
\begin{tabular}{lrrr @{\hspace{0.5cm}} lrrr}
\hline\hline
\multicolumn{4}{c}{YGG} & \multicolumn{4}{c}{YAG}\\
\hline
\multicolumn{8}{c}{unit cell parameters (\AA)}\\
$a,b,c$               &12.280&12.280&12.280&$a,b,c$              &12.008&12.008&12.008\\
$\alpha,\beta,\gamma$ &$90^\circ$&$90^\circ$&$90^\circ$&$\alpha,\beta,\gamma$&$90^\circ$&$90^\circ$&    $90^\circ$\\
\hline
\multicolumn{8}{c}{space group}\\
\multicolumn{4}{c}{$Ia\overline{3}d$ (230 setting 1)} & \multicolumn{4}{c}{$Ia\overline{3}d$ (230 setting 1)}\\
\hline
\multicolumn{8}{c}{atomic positions}\\
Y  & 0.1250 & 0.0000 & 0.2500 & Y  & 0.1250 & 0.0000 & 0.2500\\
Ga & 0.0000 & 0.0000 & 0.0000 & Al & 0.0000 & 0.0000 & 0.0000\\
Ga & 0.3750 & 0.0000 & 0.2500 & Al & 0.3750 & 0.0000 & 0.2500\\
O  & 0.0272 & 0.0558 & 0.6501 & O  & 0.9701 & 0.0506 & 0.1488\\
\hline\hline
\end{tabular}
\caption{Structural data for YGG \cite{YGGstr} and YAG \cite{YAGstr}.}
\label{strdata}
\end{table}
RE$^{3+}$ doping ions replace Y$^{3+}$ ions and hence enter the garnet structure in the dodecahedral $24c$ 
sites with the local D$_2$ symmetry. In this case each RE$^{3+}$ ion is surrounded by eight oxygen O$^{2-}$ 
ions in the dodecahedron configuration resembling a distorted cube (see Fig.~\ref{garn}). There are 24 
such sites per unit cell: half of them have absolutely identical environment with the other half;
the remaining 12 can be divided into 6 pairs where the sites differ only by inversion, and these 6 pairs differ with each other by finite rotations.
In the 
present paper we perform  calculations for the case of one particular site orientation; the coordinates of the oxygen 
atoms around the central impurity ion for that instance are presented in Table \ref{ioncoord}. 
After that, 
the results for all other sites in the unit cell can be found by applying the inversion of coordinates or the 
necessary rotations, listed in Table \ref{euler}.
\begin{table}[htb]
\begin{tabular}{l r r r r r r}
\hline
\hline
& \multicolumn{3}{c}{YGG} & \multicolumn{3}{c}{YAG} \\
\hline
& \multicolumn{1}{c}{x} & \multicolumn{1}{c}{y} & \multicolumn{1}{c}{z}
& \multicolumn{1}{c}{x} & \multicolumn{1}{c}{y} & \multicolumn{1}{c}{z} \\
\hline
O1 &  1.8690 &   0.6852 &  -1.2268 &  1.8600 &   0.6076 &  -1.2152 \\
O2 &  1.8690 &  -0.6852 &   1.2268 &  1.8600 &  -0.6076 &   1.2152 \\
O3 & -1.8690 &  -1.2268 &   0.6852 & -1.8600 &  -1.2152 &   0.6076 \\
O4 & -1.8690 &   1.2268 &  -0.6852 & -1.8600 &   1.2152 &  -0.6076 \\
O5 &  0.3082 &   2.3848 &   0.3340 &  0.2858 &   2.3944 &   0.3590 \\
O6 & -0.3082 &   0.3340 &   2.3848 & -0.2858 &   0.3590 &   2.3944 \\
O7 &  0.3082 &  -2.3848 &  -0.3340 &  0.2858 &  -2.3944 &  -0.3590 \\
O8 & -0.3082 &  -0.3340 &  -2.3848 & -0.2858 &  -0.3590 &  -2.3944 \\
\hline \hline
\end{tabular}
\caption{Coordinates of oxygen ions in YGG and YAG (\textrm{\AA})
with respect to the rare earth ion.
The axes $x$, $y$, and $z$ are directed along the three orthogonal cube edges $a$, $b$, and $c$,
Table \ref{strdata}.}
\label{ioncoord}
\end{table}

\begin{table}[htb]
\begin{tabular}{@{\hspace{0.3cm}} c @{\hspace{0.3cm}}|@{\hspace{0.7cm}} c @{\hspace{0.6cm}} c @{\hspace{0.6cm}} c 
@{\hspace{0.5cm}} c @{\hspace{0.4cm}} c @{\hspace{0.4cm}} c @{\hspace{0.4cm}}}
\hline\hline
Euler & \multicolumn{6}{c}{\hspace{-0.3cm}RE$^{3+}$ site} \\
\hspace{0.2cm}angle & 1 & 2 & 3 & 4 & 5 & 6 \\
\hline
$\alpha$&0&$\pi/2$&$\pi$&$3\pi/2$& 0     &$\pi$  \\
$\beta$ &0& 0     & 0   & 0      &$\pi/2$&$\pi/2$\\
$\gamma$&0& 0     & 0   & 0      & 0     & 0     \\
\hline\hline
\end{tabular}
\caption{
Euler angles of rotation between inequivalent RE$^{3+}$ impurity sites.}
\label{euler}
\end{table}

\section{Calculation of the electronic structure of Re$\mathbf{\cdot}$O$_{\mathbf{8}}$ cluster}

We describe an isolated impurity ion with the effective potential in the following parametric form:
\begin{eqnarray}
V_{RE}(r)&=&\frac{1}{r}\frac{(Z_i-Z)
(e^{-\frac{\mu}{d}}+1)}
{(1+\eta r)^2(e^{\frac{r-\mu}{d}}+1)}-\frac{Z_i}{r},\label{pot}\\
\textrm{Pr} &:&\mu =1.0,\  d=1.3,\  \eta =2.25 ;\nonumber\\
\textrm{Tm} &:&\mu =1.0,\  d=1.0,\  \eta =2.56.\nonumber
\end{eqnarray}
Here  $Z$ is the nuclear charge of the impurity ion, $Z_i$ is the charge of
the electron core of ion, and $\mu$, $d$ and $\eta$ are parameters that 
describe the core. We use atomic units, expressing energy in units of $E_0=27.2\ \textrm{eV}$ and
distance in units of the Bohr radius $a_B=0.53\times 10^{-8}\ \textrm{cm}$.
Solution of the Dirac equation with the potential (\ref{pot})
gives wave functions and energies of the single-electron states. The potential (\ref{pot}) provides a good 
fit to the experimental energy levels of isolated impurity ions \cite{NIST}; the comparison is presented
in Table  \ref{tab1}.

\begin{table}[h!]
\centering
\begin{tabular}{ l  l r | c r }
\hline \hline
 Ion & \multicolumn{2}{c|}{Experiment} & \multicolumn{2}{c}{Calculation} \\
&  state & energy & state & energy \\
\hline
Pr$^{2+}$ \hspace{0.5cm}& 4f$^{2}$($^3$H$_4$)5d & -155 & 5d & -153 \\
& 4f$^{2}$($^3$H$_4$)6s & -146 & 6s & -146 \\
& 4f$^{2}$($^3$H$_4$)6p & -114 & 6p & -114 \\
\hline
Pr$^{3+}$ & 4f$^{2}$($^3$H$_4$) & -314 & 4f & -313 \\
\hline\hline
\end{tabular}
\hspace{1cm}
\begin{tabular}{ l  l r | c r }
\hline \hline
 Ion & \multicolumn{2}{c|}{Experiment} & \multicolumn{2}{c}{Calculation} \\
&  state & energy & state & energy \\
\hline
Tm$^{2+}$ \hspace{0.5cm}& 4f$^{12}$($^3$H$_6$)5d & -163 & 5d & -163 \\
& 4f$^{12}$($^3$H$_6$)6s & -165 & 6s & -167 \\
& 4f$^{12}$($^3$H$_6$)6p & -126 & 6p & -126 \\
\hline
Tm$^{3+}$ & 4f$^{12}$($^3$H$_6$) & -344 & 4f & -345 \\
\hline\hline
\end{tabular}
\caption{
Calculated and experimental \cite{NIST} energy levels of an isolated ion with respect to the 
ionization limit. Energy levels are averaged over the fine structure. Units $10^3\ \textrm{cm}^{-1}\!$.}
\label{tab1}
\end{table}

In order to model the electronic structure of the RE$\cdot$O$_8$ cluster (Fig.~\ref{garn}), following 
\cite{MDS} we use the jelly model and smear the 8 oxygen ions over a spherical shell around the rare
earth ion. Hence, the effective potential due to the oxygen ions at the RE$^{3+}$ site is
\begin{equation}
V_{O}(r)=-A_o e^{-\left(\frac{r-r_o}{D}\right)^2}\!,
\label{pot1}
\end{equation}
where $r_o=4.5\ a_B$ is the mean RE--O distance, $A_o$ and $D$ are parameters of the effective potential.
To describe the electrons which contribute to the effect we use the combined spherically symmetric potential
\begin{equation}
\label{pot2}
V(\boldsymbol{r})=V_{RE}(\boldsymbol{r})+V_O(\boldsymbol{r}),
\end{equation}
where $V_{RE}$ is the single impurity ion  potential (\ref{pot}).
Solution of the Dirac equation with potential (\ref{pot2}) gives the single-particle orbitals. In this picture we 
describe the electronic configuration of the cluster as $[\textrm{RE}^{3+}]6s^26p^6$, where the electronic 
configuration of Pr$^{3+}$ is
$1s^2\! ...\,5s^25p^64f^2$ and Tm$^{3+}$ is
$1s^2\! ...\,5s^25p^64f^{12}$.
The eight states $6s^26p^6$ represent $2p_{\sigma}$-electrons of oxygens combined to $S$- and  $P$-waves
with respect to the central impurity ion (see Ref.~\cite{MDS}).
Parameter $A_o$ in the ``oxygen'' potential $V_O$ (\ref{pot1}) is determined by matching
the wavefunction of oxygen $2p_{\sigma}$-orbital (calculated in Ref.~\cite{FS}) 
with the $6s$- and $6p$-orbitals from the combined potential (\ref{pot2}) at the radius $R\approx 2.5 a_B$.
The matching conditions are
\begin{eqnarray}
\label{dual}
&&|\psi_{6s}(R)|=|\psi_{2p_{\sigma}}(r_o-R,\cos\theta=1)|,\nonumber\\
&&|\psi_{6p}(R,\cos\theta=1/\sqrt{3})|=
|\psi_{2p_{\sigma}}(r_o-R,\cos\theta=1)|,
\end{eqnarray}
This is a formulation of the idea of dual description at $r \approx R$, see Refs.~\cite{FS,Simon}.

The parameter $D$ in (\ref{pot1}) represents the size of the oxygen core and is 
about $D\lesssim 1$ (atomic units). The jelly model is rather crude and the value of $D$ cannot be 
determined precisely, see Ref.~\cite{MDS}.
In the present work we vary this parameter in the range of $0.1$--$1.5\,$. For each particular value of
$D$ we find $A_o$ to satisfy (\ref{dual}), for example $A_o=0.9$ at $D=1\,$. The most realistic value for
$D$ is probably around $0.5$--$1.0\,$. To be specific, in the final answers we present results at $D=1.0\,$.
Instead of the jelly model it would certainly be better to use a relativistic quantum chemistry 
Hartree-Fock method \cite{GQ} (or the  Kohn-Sham form of the relativistic density functional method which allows one
to generate electron orbitals) to describe the RE$\cdot$O$_8$ cluster.
However, this would be a much more involved calculation at the edge of present computational
capabilities and therefore, at this stage, we continue with the jelly model.

\section{Calculation of the effective Hamiltonian (\ref{HdynI})}

The calculations in the present section are similar to those performed in \cite{MDS} for the electric
dipole moment of the electron.
There are three perturbation  operators that contribute to the correlation (\ref{HdynI}).
First, there is a magnetic interaction of the electron with the nuclear anapole moment, see, e.g., \cite{Khripl}.
Expressed in atomic units the interaction reads
\begin{eqnarray}
\label{pert1}
V_a&=&K_a
(\boldsymbol{I}\boldsymbol{\alpha}) \delta(\boldsymbol{r}) ,\\
K_a&=&S_a \kappa_a \left( \frac{Gm^2\alpha}{\sqrt{2}}\right)= 1.57\cdot 10^{-14} \kappa_a S_a,\nonumber\\
\kappa_a &=&
\frac{9}{10}g\frac{\alpha \mu}{m \tilde{r}_0}A^{2/3};\nonumber\\
^{141}\textrm{Pr} : \kappa_a &=& 0.35, \ \ \ \ S_a=-0.34,\nonumber\\
^{169}\textrm{Tm} : \kappa_a &=& 0.39, \ \ \ \ S_a=-0.25 .\nonumber
\end{eqnarray}
Here $m$ is the electron mass, $G$ is the Fermi constant, and $\alpha$ is the fine structure constant;
$\boldsymbol{\alpha}$ are the Dirac matrices, $\mu$ is the magnetic moment of the unpaired nucleon (proton in
these cases) expressed in nuclear magnetons, $\tilde{r}_0 =1.2fm$, $A$ is the mass number of the nucleus, and 
$g\approx4$ for outer proton and  $g\sim1$ for outer neutron. Values of the nuclear structure constant 
$S_a$ have been calculated in \cite{MS}.

\begin{figure}[!hbt]
\centering
\epsfig{figure=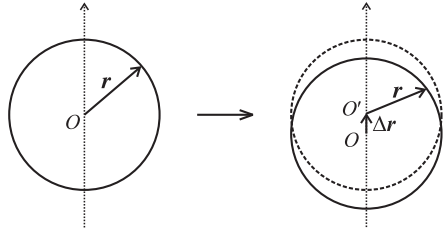,height=3cm}
\caption{Schematic picture, illustrating the shift of ${V_O(\boldsymbol{r})}$
due to the lattice deformation.}
\label{Figshift}
\end{figure}

The second perturbation operator is related to the shift $\Delta\boldsymbol{r}$ of the rare earth ion with 
respect to the surrounding oxygen ions.  The shift is proportional to the external electric field, 
but for now we consider $\Delta\boldsymbol{r}$ as an independent variable.
In the jelly model $-\Delta\boldsymbol{r}$ is the shift of the spherically symmetric oxygen potential $V_O(r)$ 
(\ref{pot1}) with respect to the origin, see Fig.~\ref{Figshift}.
Therefore,
\begin{eqnarray}
V_O(\boldsymbol{r}) \rightarrow V^{'} _O(\boldsymbol{r})
&=&V_O(\boldsymbol{r}+ 
\Delta\boldsymbol{r})= V_O(\boldsymbol{r})+\frac{(\Delta\boldsymbol{r\cdot r)}}{r}
\frac{\partial V_O}{\partial r} 
\end{eqnarray}
Thus, the perturbation operator related to the lattice deformation reads
\begin{eqnarray}
V_1(\boldsymbol{r})&=&\frac{(\Delta\boldsymbol{r\cdot r)}}{r}
\frac{\partial V_O}{\partial r} 
=\left(\Delta x \sin\theta \cos\phi + \Delta y \sin\theta\sin\phi + \Delta z \cos\theta\right) 
(-2)\frac{(r-r_o)}{D^2}V_O(r) .
\label{pert2}
\end{eqnarray}
Here $\boldsymbol{r}=r(\sin\theta \cos\phi, \sin\theta\sin\phi, \cos\theta)$.

The third perturbation is the residual electron-electron Coulomb interaction, which is not included in the
effective potential,
\begin{equation}
\label{pert3}
V_C(\boldsymbol{r}_i,\boldsymbol{r}_j)=\frac{1}{|\boldsymbol{r}_i -
\boldsymbol{r}_j|}=
\sum_{lm}\frac{4\pi}{2l+1}\frac{r_<^l}{r_>^{l+1}}
Y^*_{lm}(\boldsymbol{r}_i) Y_{lm}(\boldsymbol{r}_j).
\end{equation}
Here $\boldsymbol{r}_i$ and $\boldsymbol{r}_j$ are radius-vectors of the two interacting 
electrons.

The formula for the energy correction in the third order of perturbation theory reads, see, e.g., Ref.~\cite{LL}:
\begin{equation}
E_n^{(3)}=\sum_m {}^{'}\sum_k {}^{'}\frac{V_{nm}V_{mk}V_{kn}}{\hbar^2 
\omega_{mn} \omega_{kn}} - V_{nn}\sum_m {}^{'}\frac{|V_{nm}|^2}{\hbar^2 
\omega_{nm}^2},
\label{3d}
\end{equation}
where $V=V_a+V_1+V_C$. In Eq.~(\ref{3d}) we need to consider only the terms that contain all
the operators $V_a$, $V_1$, and $V_C$.

\begin{figure}[h!]

\hspace{-5.0cm}
\begin{tabular}{c c c c}

\multicolumn{1}{m{1cm}}{$-\quad 2 \times$} &
\multicolumn{1}{m{3.6cm}}{\epsfig{figure=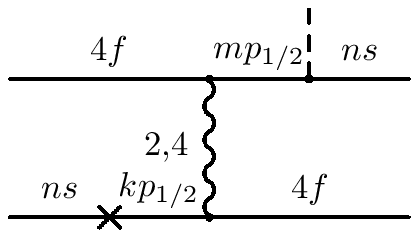,height=2cm}}&
\multicolumn{1}{m{1cm}}{$-\quad 2 \times$} &
\multicolumn{1}{m{3.6cm}}{\epsfig{figure=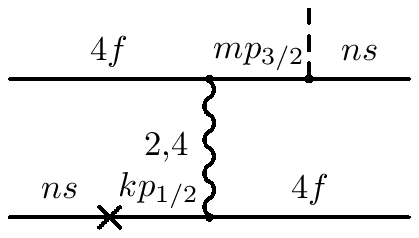,height=2cm}}\\
&1)& &2) \\

\multicolumn{1}{m{1cm}}{$-\quad 2 \times$} &
\multicolumn{1}{m{3.6cm}}{\epsfig{figure=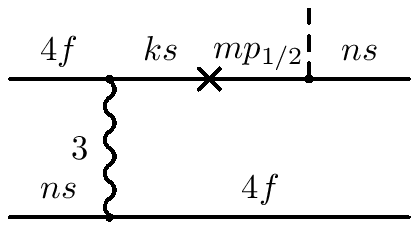,height=2cm}}&
\multicolumn{1}{m{1cm}}{$-\quad 2 \times$} &
\multicolumn{1}{m{3.6cm}}{\epsfig{figure=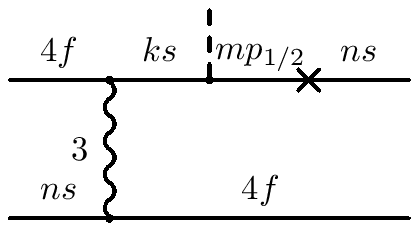,height=2cm}}\\
&3)& &4)
\end{tabular}

\vspace{0.2cm}
\hspace{-3.0cm}
\begin{tabular}{l c c c c c c}

\multicolumn{1}{m{0.8cm}}{$+$} &
\multicolumn{1}{m{1.8cm}}{\epsfig{figure=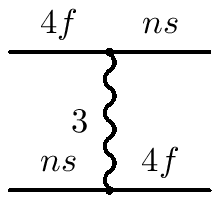,height=1.6cm}}&
\multicolumn{1}{m{0.8cm}}{$\times\quad\!\!\Bigg($} &
\multicolumn{1}{m{2.7cm}}{\epsfig{figure=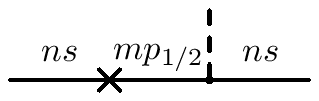,height=0.76cm}}&
\multicolumn{1}{m{0.7cm}}{$\quad\!\!\!+$} &
\multicolumn{1}{m{2.7cm}}{\epsfig{figure=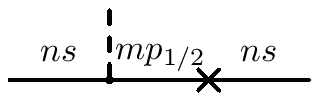,height=0.76cm}}&
\multicolumn{1}{m{1cm}}{$\Bigg)$}
\\& &5)& & & &
\end{tabular}

\vspace{0.1cm}

\begin{tabular}{c c c c c c}

\multicolumn{1}{m{1cm}}{$-\quad 2 \times$} &
\multicolumn{1}{m{3.6cm}}{\epsfig{figure=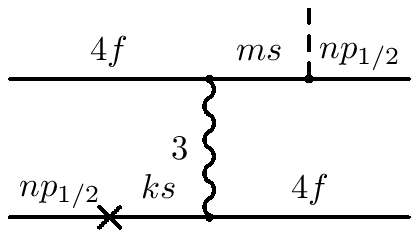,height=2cm}}&
\multicolumn{1}{m{1cm}}{$-\quad 2 \times$} &
\multicolumn{1}{m{3.6cm}}{\epsfig{figure=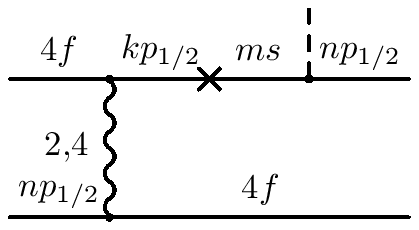,height=2cm}}&
\multicolumn{1}{m{1cm}}{$-\quad 2 \times$} &
\multicolumn{1}{m{3.6cm}}{\epsfig{figure=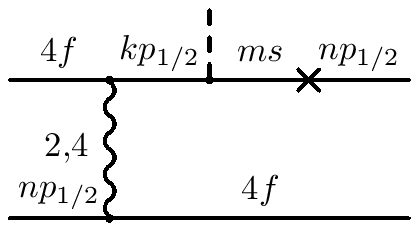,height=2cm}}\\
&6)& &7)& &8)\\

\multicolumn{1}{m{1cm}}{$-\quad 2 \times$} &
\multicolumn{1}{m{3.6cm}}{\epsfig{figure=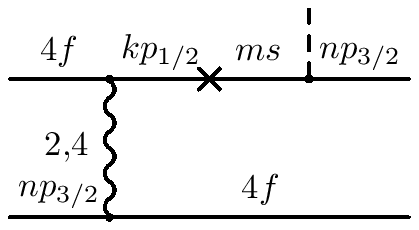,height=2cm}}&
\multicolumn{1}{m{1cm}}{$-\quad 2 \times$} &
\multicolumn{1}{m{3.6cm}}{\epsfig{figure=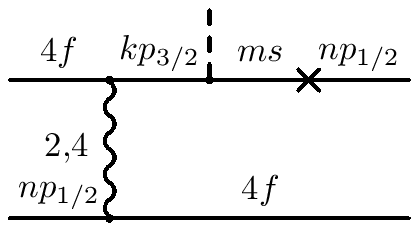,height=2cm}}&
&\\
&9)& &10)& &
\end{tabular}

\vspace{0.2cm}
\hspace{-3.0cm}
\begin{tabular}{l c c c c c c}

\multicolumn{1}{m{0.8cm}}{$+$} &
\multicolumn{1}{m{1.8cm}}{\epsfig{figure=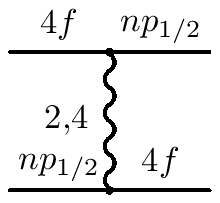,height=1.6cm}}&
\multicolumn{1}{m{0.8cm}}{$\times\quad\!\!\Bigg($} &
\multicolumn{1}{m{2.7cm}}{\epsfig{figure=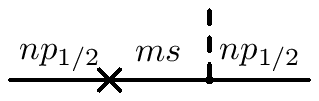,height=0.76cm}}&
\multicolumn{1}{m{0.7cm}}{$\quad\!\!\!+$} &
\multicolumn{1}{m{2.7cm}}{\epsfig{figure=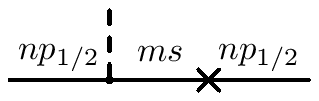,height=0.76cm}}&
\multicolumn{1}{m{1cm}}{$\Bigg)$}
\\
& &11)& & & &
\end{tabular}
\caption{Third order perturbation theory diagrams corresponding to
Eq.~(\ref{3d}). The cross denotes the anapole interaction ${V_a}$ (\ref{pert1}),
the dashed line denotes the lattice deformation perturbation ${V_1}$ (\ref{pert2}),
and the wavy line denotes the Coulomb interaction ${V_C}$ (\ref{pert3}). The multipolarity
of the Coulomb interaction is shown near the wavy line.
Each diagram contributes with the coefficient shown before the diagram (number of diagrams of this kind).
Summation over {\it all} intermediate states ${|k\rangle}$ and ${|m\rangle}$ and
over {\it all filled} states ${|n\rangle}$ is assumed.}
\label{alld}
\end{figure}

The shift operator $V_1$ is nearly saturated by $6s$- and $6p$-states because core electrons do not
``see'' the deformation of the lattice, hence, for this operator we consider only $s$-$p$ mixing.
Matrix elements of the anapole operator $V_a$ practically vanish for the electron states with 
high angular momentum, since this operator is proportional to the Dirac delta function. 
Therefore, it is sufficient to take into account only $\langle ns_{1/2}|V_1| kp_{1/2}\rangle$ matrix elements.
All in all, there are 11 diagrams (Fig.~\ref{alld}) that correspond to Eq.~(\ref{3d}). All diagrams are 
exchange ones and contribute with the sign shown before each of the diagrams. Summation over {\it all} 
intermediate states $|k\rangle$ and $|m\rangle$ and over {\it all filled} states $|n\rangle$ is assumed.

Since $V_a$ and $V_1$ are single-particle operators, we evaluate each diagram by
solving equations for the corresponding wavefunction corrections. For example, the first diagram contains 
in the top right leg the correction
\begin{equation}
\label{dp1}
|\delta\psi_x\rangle = \sum_m \frac{\langle mp_{1/2}|V_1|ns\rangle}
{\epsilon_{ns}-\epsilon_{mp_{1/2}}}|mp_{1/2}\rangle .
\end{equation}
To evaluate the correction we do not use a direct summation, but instead
solve the equation
\begin{equation}
\label{dp2}
(H-\epsilon)|\delta\psi_x\rangle = -V_1|ns\rangle , \ \ \ \epsilon =\epsilon_{ns}
\end{equation}
for each particular $|ns\rangle$ state. Here $H$ is the Dirac Hamiltonian
with the potential (\ref{pot2}). Similarly, the bottom left leg of the same diagram is evaluated using
\begin{equation}
\label{dp3}
(H-\epsilon_{ns})\delta\psi_d = -V_a|ns\rangle.
\end{equation}
In solving this equation we take the finite size of the nucleus into account by replacing the $\delta$-function
in (\ref{pert1}) with a realistic nuclear density.

\begin{figure}[!hbt]
\centering
\epsfig{figure=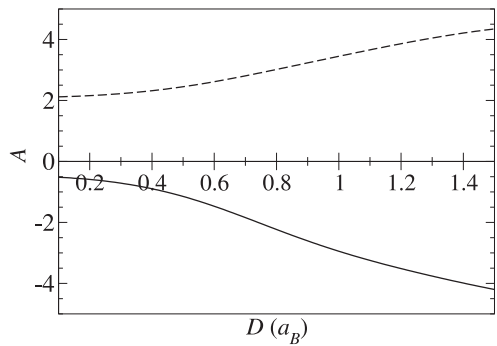,height=4.5cm}
\caption{
Value of the coefficient $A$ defined in Eq.~(\ref{de}) versus width of the effective oxygen potential. 
The dashed line corresponds to Pr$^{3+}$ in YGG and the solid line corresponds to Tm$^{3+}$ ions in YAG.
}
\label{FigDA}
\end{figure}

Apart from the coefficients presented in Fig.~\ref{alld}, which in essence show the number of diagrams of
each kind, each particular diagram in Fig.~\ref{alld} contributes with its own angular coefficient.
In calculating the coefficients we assumed, without loss of generality, that the total angular
momentum of the $4f$-electrons is directed along the $z$-axis, $|J,J_z\rangle$.
Values of the coefficients are presented in Table \ref{ang} in the Appendix.
The method for separating the radial equations corresponding to (\ref{dp2}) and (\ref{dp3})
is also described in the Appendix.
As the result of the calculations we find the following $P$-odd energy correction
related to the displacement $\Delta \boldsymbol{r}$ of the RE impurity ion:
\begin{equation}
\Delta \epsilon = K_a \alpha A \frac{1}{a_B}
(\boldsymbol{\Delta r}\cdot[\boldsymbol{I}\times\boldsymbol{J}])E_0 .
\label{de}
\end{equation}
We recall that $\boldsymbol{I}$ is spin of the nucleus, $\boldsymbol{J}$ is the total angular momentum 
of the $f$-electrons, $E_0=27.2\ \textrm{eV}$ is the atomic unit of energy, $a_B$ is the Bohr radius, $\alpha$ is the fine
structure constant, and $K_a$ is given in Eq.~(\ref{pert1}).
The dimensionless coefficient $A$ for the Pr$^{3+}$ and Tm$^{3+}$ ions (in the corresponding lattices) calculated at $D=1.0$ in 
Eq.~(\ref{pot1}) reads:
\begin{eqnarray}
A_{\mathrm{Pr}}&=&
 -25.99
 -11.20
  +0.32
  +0.59
  +18.64
 -18.99\nonumber \\&&{}
   +0.58
   +1.39
 -15.99
  +28.37
  +25.73=3.45 , \nonumber
\\
{}\nonumber\\
A_{\mathrm{Tm}}&=&
 9.77
+ 12.78
 -3.58
 -1.33
-32.24
+ 36.49\nonumber \\&&{}
+  0.21
+  0.12
+ 53.48
-67.70
-10.95=   -2.95 .
\label{ATm}
\end{eqnarray}
The eleven  terms in (\ref{ATm}) represent the contributions of the eleven diagrams in Fig.~\ref{alld}.
As one can see, there is significant compensation between different terms in (\ref{ATm}).
This compensation is partially related to the fact that each particular diagram in
Fig.~\ref{alld} contains contributions forbidden by the Pauli principle.
These contributions are canceled out only in the sum of the diagrams.
To check (\ref{ATm}) we have also performed a more involved calculation 
explicitly taking into account the Pauli principle in each particular diagram, the results read:
\begin{eqnarray}
\label{APauli}
A_{\mathrm{Pr}} &=&
  -0.07
  -0.18
+   1.48
+   0.72
+  1.00
  -2.63\nonumber \\&&{}
  -1.74
+   0.88
+  41.34
 -40.00
+   2.65=    3.45 , \nonumber\\
{}\nonumber \\
A_{\mathrm{Tm}} &=&
  -0.55
  -0.56
  -4.09
  -1.36
  -1.47
+   6.27\nonumber \\&&{}
+   0.53
+   0.15
 -38.94
+  38.06
  -0.99=   -2.95 .
\label{ATm1}
\end{eqnarray}
Although each individual term has changed compared to  (\ref{ATm}), the total sum of the diagrams remains the same.
Comparison between (\ref{ATm}) and (\ref{ATm1}) is a test of the many body perturbation theory used in
the calculation. To demonstrate the sensitivity to parameters of the effective potential,
we plot in Fig.~\ref{FigDA} the coefficient $A$ versus the width $D$ of the oxygen potential, see Eq.~(\ref{pot1}).
As we pointed out in Section III, the most realistic value of $D$ is around $0.5$--$1.0\,$. 
To be specific, in the final estimates we use the results (\ref{ATm}) and (\ref{ATm1}), which correspond to the value $D=1.0\,$.

\section{Calculation of the effective Hamiltonian (\ref{HstI})}
\label{subsB}

The  $P$-odd effective Hamiltonian considered in the previous section arises due to a
shift of the environment with respect to the rare earth ion. In other words, it is
due to the first harmonic in the electron density induced by the perturbation operator $V_1$ (\ref{pert2}). 
In the equilibrium position the first harmonic vanishes identically due to the symmetry of the lattice. 
The next harmonic in the electron density that contributes to the parity nonconserving effect is the third 
harmonic which is  nonzero even in the equilibrium position of the rare earth ion. This effect gives  
the $P$-odd energy shift even in the absence of an external electric field.

The effective oxygen potential $V_O$ (\ref{pot1}) represents the spherically symmetric part of the real 
potential for electrons created by the eight oxygen ions in the garnet lattice. Let us describe the potential
(pseudopotential) of a single oxygen ion as $g \delta(\boldsymbol{r}-\boldsymbol{R})$, where
$\boldsymbol{R}$ is the position of the ion and $g$ is some constant. Then the total potential is
\begin{equation}
V(\boldsymbol{r})=\sum_{\boldsymbol{R}} g \delta(\boldsymbol{r}-\boldsymbol{R}),
\label{V}
\end{equation}
where  summation is performed over the coordinates of the eight oxygen ions presented in Table \ref{ioncoord}.
Expanding the Dirac delta function in the potential $V(\boldsymbol{r})$ in a series of spherical 
harmonics, we find
\begin{equation}
V(\boldsymbol{r})=g\frac{\delta(r-R)}{R^2}\sum_{km}\sum_{\boldsymbol{R}}
Y^*_{km}(\boldsymbol{R})\cdot Y_{km}(\boldsymbol{r}).
\end{equation}
Then,
\begin{eqnarray}
V_O(\boldsymbol{r})&=&g\frac{\delta(r-R)}{R^2}\sum_{\boldsymbol{R}}
Y_{00}(\boldsymbol{R})\cdot Y_{00}(\boldsymbol{r})\to-A_o e^{-\left(\frac{r-r_o}{D}\right)^2}\!,
\end{eqnarray}
and hence the third harmonic reads
\begin{eqnarray}
V_{3}(\boldsymbol{r})&=&g\frac{\delta(r-R)}{R^2}
\sum_{\boldsymbol{R}}Y^*_{3m}(\boldsymbol{R})\cdot Y_{3m}(\boldsymbol{r})
\to-A_o e^{-\left(\frac{r-r_o}{D}\right)^2} \frac{\pi}{2} T_{3m}\cdot Y_{3m}(\boldsymbol{r}),\label{pert3m}\\
T_{3m}&=&\sum_{\boldsymbol{R}}Y^*_{3m}(\boldsymbol{R}) .\nonumber
\label{t3m}
\end{eqnarray}
The spherical tensor $T_{3m}$ (lattice octupole) for yttrium aluminium garnet and yttrium gallium garnet
has only one non-zero independent component, $T_{31}=-0.1876$ for YAG and $T_{31}=-0.1010$ for YGG.
All other components are determined by the following relations:
\begin{equation}
T_{33}=\sqrt{\frac{3}{5}}T_{31},\ \ T_{3\,\textrm{-}1}=-T_{31},\ \ T_{3\,\textrm{-}3}=-T_{33},
\ \ T_{30}=0.
\end{equation}
Components of the corresponding Cartesian  irreducible tensor  $T_{klm}$ can be found using the following
relations:
\begin{equation}
T_{xzz}=T_{zxz}=T_{zzx}=-T_{xyy}=-T_{yxy}=-T_{yyx}=-\sqrt{\frac{8}{15}}T_{31} .
\end{equation}
All other components of the Cartesian tensor are equal to zero.

\begin{figure}[h!]
\begin{tabular}{c c c c}

\multicolumn{1}{m{1cm}}{$-\quad 2 \times$} &
\multicolumn{1}{m{3.6cm}}{\epsfig{figure=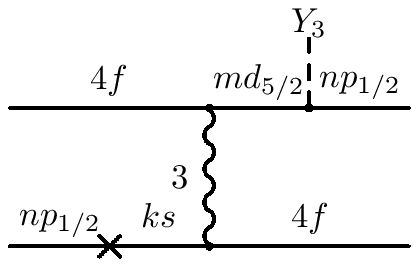,height=2cm}}&
\multicolumn{1}{m{1cm}}{$-\quad 2 \times$} &
\multicolumn{1}{m{3.6cm}}{\epsfig{figure=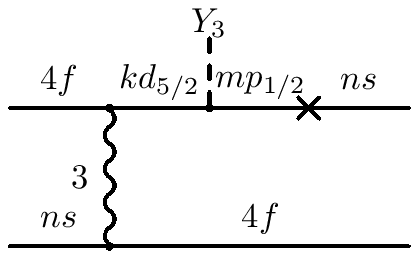,height=2cm}}\\
&1)& &2) \\

\multicolumn{1}{m{1cm}}{$-\quad 2 \times$} &
\multicolumn{1}{m{3.6cm}}{\epsfig{figure=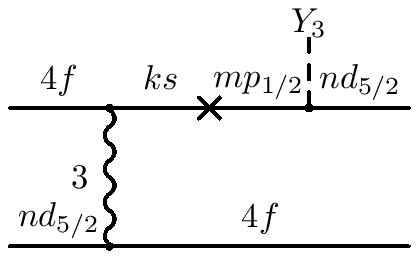,height=2cm}}&&\\
&3)& &\\

\multicolumn{1}{m{1cm}}{$-\quad 2 \times$} &
\multicolumn{1}{m{3.6cm}}{\epsfig{figure=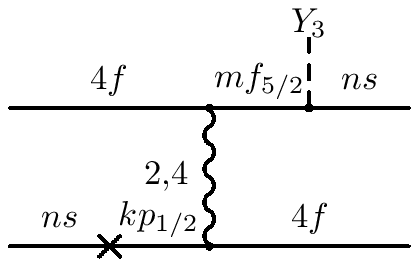,height=2cm}}&
\multicolumn{1}{m{1cm}}{$-\quad 2 \times$} &
\multicolumn{1}{m{3.6cm}}{\epsfig{figure=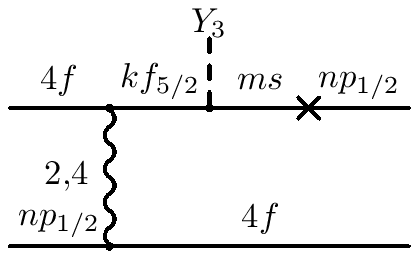,height=2cm}}\\
&4)& &5)\\

\multicolumn{1}{m{1cm}}{$-\quad 2 \times$} &
\multicolumn{1}{m{3.6cm}}{\epsfig{figure=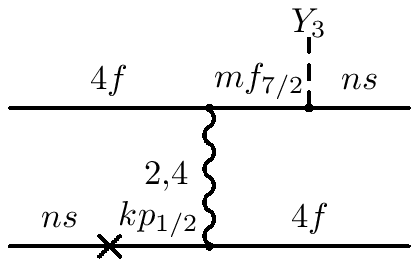,height=2cm}}&
\multicolumn{1}{m{1cm}}{$-\quad 2 \times$} &
\multicolumn{1}{m{3.6cm}}{\epsfig{figure=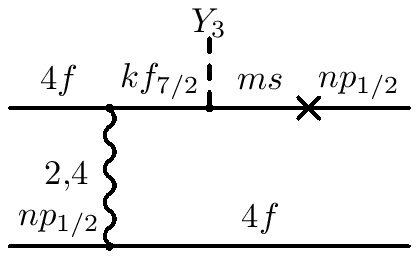,height=2cm}}\\
&6)& &7)
\end{tabular}
\caption{Diagrams for the ``octupole'' effect.
The cross denotes the anapole interaction ${V_a}$ (\ref{pert1}),
the dashed line denotes the lattice octupole ${V_3}$ (\ref{t3m}),
and the wavy line denotes the Coulomb interaction ${V_C}$ (\ref{pert3}). 
The multipolarity of the Coulomb interaction is shown near the wavy line.
Each diagram contributes with the coefficient shown before the diagram
 (number of diagrams of this kind).
Summation over {\it all} intermediate states ${|k\rangle}$ and ${|m\rangle}$ and
over {\it all filled} states ${|n\rangle}$ is assumed.}
\label{Y3diags}
\end{figure}

Similar to the ``dipole'' effect considered in the previous section, the octupole effect arises in the 
third order of perturbation theory.  The relevant perturbation theory operators are 
a) interaction of the electron with the nuclear anapole moment $V_a$ (\ref{pert1}), 
b) interaction of the electron with the lattice octupole harmonic $V_{3}$ (\ref{pert3m}), and 
c) the residual electron-electron Coulomb interaction $V_C$ (\ref{pert3}).
The formula for the energy correction (\ref{3d}) yields 7 diagrams which are presented in Fig.~\ref{Y3diags}.

Besides the coefficients presented in Fig.~\ref{Y3diags}, which show the number of diagrams of
each kind, each particular diagram in Fig.~\ref{Y3diags} contributes with its own angular coefficient.
In calculating the coefficients we assumed, without loss of generality, that the total angular
momentum of $4f$ electrons is directed along the $z$-axis, $|J,J_z\rangle$, and
the nuclear spin is directed along the $y$-axis, $\boldsymbol{I}=(0,I,0)$.
The angular coefficients for each of the 7 diagrams from Fig.~\ref{Y3diags} are presented in 
Table \ref{ang} in the Appendix. The method for separating the radial equations is also described in the Appendix.
The effective Hamiltonian for the lattice octupole effect has the following form
\begin{equation}
\Delta \epsilon =K_a\alpha B I_i \epsilon_{ijk} T_{klm} \left( J_j J_l J_m + J_m J_l J_j \right)E_0 .
\label{effHam}
\end{equation}
Eq.~(\ref{effHam}) represents the only $P$-odd scalar combination one can construct from the two 
vectors and one irreducible 
third rank tensor. Note, that $\boldsymbol{J}$ here is an operator, and different components of
$\boldsymbol{J}$ do not commute. This is why in the right hand side of Eq.~(\ref{effHam}) we
explicitly write the Hermitian combination. The matrix element of (\ref{effHam}) in the kinematics
which we consider for the calculation of the angular coefficients (Table \ref{ang}) is
\begin{equation}
\langle J,J_z| I_i \epsilon_{ijk} T_{klm} \left( J_j J_l J_m + J_m J_l J_j \right)|J, J_z\rangle=
T_{zzx}IJ_z[5J_z^2-3J(J+1)+1] .
\end{equation}

\begin{figure}[!hbt]
\centering
\epsfig{figure=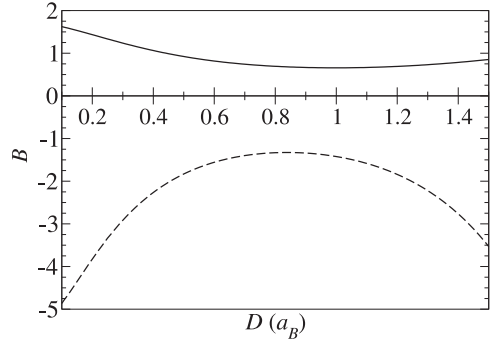,height=4.5cm}
\caption{
Value of the coefficient $B$ in Eq.~(\ref{effHam}) versus the width of the effective oxygen potential. 
Dashed line corresponds to Pr$^{3+}$ in YGG and solid line corresponds to Tm$^{3+}$ in YAG.
}
\label{DAY3}
\end{figure}

Our calculations show that contributions of the diagrams with the intermediate $f$-state 
(diagrams 4,5,6,7 in Fig.~\ref{Y3diags}) are at least 30 times smaller compared to diagrams 1 and 2.
The reason for this is very simple: $f$-electrons are practically decoupled from the lattice
deformation. The diagram 3 is even smaller because internal $3d$- and $4d$-electrons are also
decoupled from the lattice. So, only diagrams 1 and 2 contribute to the effect and they are
nearly saturated by the intermediate unoccupied $5d$-state. 
The dimensionless coefficient $B$ for Pr and Tm ions in corresponding lattices
calculated at $D=1.0$ [Eq.~(\ref{pot1})] reads:
\begin{eqnarray}
B_{\mathrm{Pr}}(D=1)&=&
 -2.18
 +0.76 = -1.42, \nonumber\\
B_{\mathrm{Tm}}(D=1)&=&
 1.11
-0.45=0.66.  
\label{eq45}
\end{eqnarray}
The two terms in equations (\ref{eq45}) represent the contributions of the first and second 
diagrams.
The variation of the coefficient $B$ with the width of the effective oxygen potential $D$
is shown in Fig.~\ref{DAY3}.
Again, we recall that the most realistic value of $D$ is around $0.5$--$1.0\,$. 
To be specific, in the estimates for the effect we use $D=1.0\,$.  

\section{Crystal field, average electron magnetization, orientation of nuclear spin}

The energy of a free ion is degenerate with respect to the $z$-projection of total angular momentum. 
Interaction with the lattice (crystal field) breaks the rotational invariance and lifts the degeneracy.
The effective crystal-field Hamiltonian can be written in the following form, see, e.g., \cite{CFI}
\begin{equation}
H_{\textrm{\it cf}}=\sum_{km}B^*_{km}\sum_i \sqrt{\frac{4\pi}{2k+1}} Y_{km}(\boldsymbol{r})
\end{equation}
where $B_{km}$ are the crystal field parameters, $\boldsymbol{r}$ is the radius-vector of the 
atomic electron.

Experimental values of the energy levels for Pr$^{3+}$ in YGG and Tm$^{3+}$ in YAG are known \cite{Pr,Tm},
and fits of the crystal field parameters $B_{km}$ have been performed in the experimental
papers. Unfortunately, we cannot use these fits because they are performed without connection
to a particular orientation of crystallographic axes. We need to know the connection and
therefore we have performed independent fits. For the fits we use a modified
point-charge model. In the simple point-charge  model the crystal field is of the form
\begin{eqnarray}
A_{km}^{(pc)}&=&-\sum_j\frac{q_j}{r_j^{k+1}}\sqrt{\frac{4\pi}{2k+1}}Y_{km}(\boldsymbol{r}_j),\\
B_{km}^{(pc)}&=&\rho_k A_{km}^{(pc)},
\end{eqnarray}
where $j$ enumerates ions of the lattice and $\rho_k=\langle r^k \rangle$ is the expectation
value over the RE $f$-electron wave function. The values of $\rho_k$ are known \cite{CFI}.
The point charges are $q_O = -2$ and $q_Y=q_{\textrm{\it Ga}}=q_{\textrm{\it Al}}=3$.
Clearly, the naive point-charge model is insufficient to describe the nearest 8 oxygen ions because of
the relatively large size of the ions (extended electron density of the host oxygens).
To describe the effect of the extended electron density we introduce an additional
field $A_{km}^{(el)}$
\begin{eqnarray}
A_{km}&=&A_{km}^{(pc)}+A_{km}^{(el)},\\
A_{km}^{(el)}&=&-\alpha_k\sum_{j=1}^8\frac{q_j}{r_j^{n+1}}\sqrt{\frac{4\pi}{2n+1}}Y_{km}(\boldsymbol{r}_j),
\end{eqnarray}
here the sum runs over the eight oxygen ions surrounding the dopant ion in the garnet structure,
and $\alpha_k$ are fitting parameters. So, we have only three fitting parameters,
$\alpha_2$, $\alpha_4$, and $\alpha_6$, because higher multipoles do not contribute in $f$-electron
splitting. In the end, we get a fairly good fit of the experimental energy levels, see 
Table \ref{crfieldlev}. The values of the resulting crystal field parameters $B_{km}$ are presented 
in Table \ref{crfieldpar}.
\begin{table}
\begin{tabular}{ r @{\hspace{0.5cm}} r | r @{\hspace{0.5cm}} r }
\hline
\hline
\multicolumn{2}{c}{Pr$^{3+}$:YGG} & \multicolumn{2}{c}{Tm$^{3+}$:YAG} \\
\hline
Exp.~\cite{Pr} & Calc. & Exp.~\cite{Tm} & Calc. \\
\hline
0   & 0   & 0   & 0   \\
23  & 23  & 27  & 27  \\
23  & 23  & 216 & 182 \\
-   & 400 & 240 & 240 \\
532 & 413 & 247 & 253 \\
578 & 538 & 300 & 301 \\
598 & 621 & 450 & 306 \\
626 & 877 & 588 & 494 \\
689 & 895 & 610 & 609 \\
         && 650 & 673 \\
         && 690 & 686 \\
         && 730 & 825 \\
         && -   & 937 \\
\hline\hline
\end{tabular}
\caption{Experimental and calculated crystal field energy levels in $cm^{-1}$. $J$-$J$ mixing is neglected in the calculation.}
\label{crfieldlev}
\end{table}

\begin{table}
\begin{tabular}{ l | r r r r r r r r r r r r r r r}
\hline\hline
Compound & $B_{20}$ & $B_{21}$ & $B_{22}$ & $B_{40}$ & $B_{41}$ & $B_{42}$ & $B_{43}$ & $B_{44}$ & $B_{60}$ & $B_{61}$ & $B_{62}$ & $B_{63}$ & $B_{64}$ & $B_{65}$ & $B_{66}$\\
\hline
Pr:YGG & 622 & 11$i$ & -762 & 211 & -475$i$ & 727 & 1256$i$ & -423 & 963 & -280$i$ & 
-648 & -437$i$ & 91 & 304$i$ & -961\\
Tm:YAG &  257 & 92$i$ & -315 & -1198 & 344$i$ & -248 & -909$i$ & -523 & -938 & 528$i$ & 569 &
816$i$ & 94 & -563$i$ & 843 \\
\hline\hline
\end{tabular}
\caption{Crystal field parameters in cm$^{-1}$, that fit the energy levels in Table \ref{crfieldlev}.}
\label{crfieldpar}
\end{table}

\begin{figure}[htb]
\centering
\vspace{0.5cm}
\epsfig{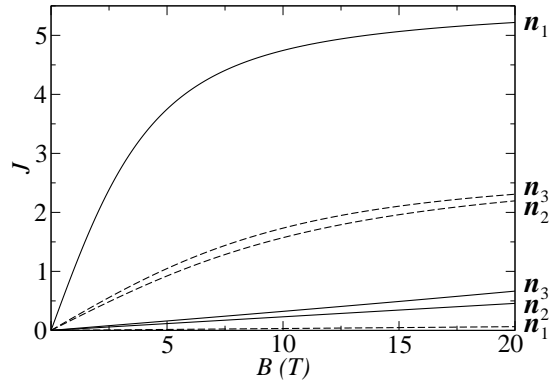}
\caption{
The average total electron angular momentum of the rare earth ion versus magnetic field (tesla).
Directions of the magnetic field correspond to the principal axes of the magnetization tensor
$\boldsymbol{n}_1$, $\boldsymbol{n}_2$, and $\boldsymbol{n}_3$. Solid lines correspond to Tm$^{3+}$
in YAG and dashed lines correspond to Pr$^{3+}$ in YGG.}
\label{magn}
\end{figure}

For the non-Kramers ions, such as Pr$^{3+}$ and Tm$^{3+}$, the expectation value of the 
total angular momentum in the ground state vanishes due to the crystal field, 
$\langle \boldsymbol{J}\rangle=0$. To get a nonzero $\langle \boldsymbol{J}\rangle$ one needs to 
apply an external magnetic field $\boldsymbol{B}$.
Diagonalizing the Hamiltonian matrix of the dopant ion in the magnetic field
\begin{equation}
\langle J^\prime_z|H_{\textrm{\it cf}} + \mu_B g (\boldsymbol{J} \boldsymbol{B}) |J_z \rangle,
\end{equation}
($9\times 9$ matrix for Pr$^{3+}$ and $13\times 13$ matrix for  Tm$^{3+}$)
we find the ground state of the ion in the presence of the external magnetic field $B$ 
(here $\mu_B$ is the Bohr magneton and $g$ is the atomic Lande factor; $g=0.80$ for  Pr$^{3+}$ 
in $^3\! H_4$ configuration and $g=1.17$ for Tm$^{3+}$ in $^3\! H_6$ configuration.) 
For weak magnetic field the average total angular momentum can be written as
\begin{equation}
\langle J_i \rangle = \tau_{ik} B_k .
\label{Jav}
\end{equation}
The tensor $\tau_{ik}$ can be diagonalized. According to our calculations,
both for Pr and Tm it is diagonal with the principal axes $\boldsymbol{n}_1=(1,0,0)$,
$\boldsymbol{n}_2=(1,1/\sqrt{2},1/\sqrt{2})$,
$\boldsymbol{n}_3=(1,1/\sqrt{2},-1/\sqrt{2})$:
\begin{eqnarray}
\textrm{Pr}: \ \ \ \ \tau=\left(
\begin{array}{ccc}
-0.003 & 0      &0\\
0 & -0.154 &0\\
0 & 0 &-0.176
\end{array}\right)\frac{1}{tesla},\ \ \ \ \ \
\textrm{Tm}: \ \ \ \ \tau=\left(
\begin{array}{ccc}
-0.474 & 0 &0\\
0 & -0.023      &0\\
0 &  0 &-0.032
\end{array}\right)\frac{1}{tesla}.
\end{eqnarray}
The average total electron angular momentum in the magnetic fields applied along the directions $\boldsymbol{n}_1$, $\boldsymbol{n}_2$, and $\boldsymbol{n}_3$ is 
plotted in Fig.~\ref{magn}. We see that the linear expansion (\ref{Jav}) is valid for the field $B < 5$--10 T.

The effective Hamiltonian for the nuclear spin is
\begin{equation}
H_{\textrm{\it nuc}}=A_{\textrm{\it hf}} (\boldsymbol{J\cdot I}) - \frac{\mu \mu_N}{I} (\boldsymbol{B\cdot I}),
\label{Hnuc}
\end{equation}
where $A_{\textrm{\it hf}}$ is the hyperfine  constant, $\mu$ is the nuclear  magnetic moment
in nuclear magnetons and $\mu_N$ is the nuclear magneton:
\begin{eqnarray}
^{141}\textrm{Pr} &:&\ A_{\textrm{\it hf}} = 1093\  \textrm{MHz}\ \textrm{\cite{ahf}},\ \mu = 4.2754\ \textrm{\cite{tabisot}},
\ I = 5/2;\nonumber\\
^{169}\textrm{Tm} &:&\ A_{\textrm{\it hf}} = -393.5\  \textrm{MHz}\ \textrm{\cite{ahf}},\ \mu = -0.2316\ \textrm{\cite{tabisot}},\ I = 1/2.
\end{eqnarray}

\begin{figure}[htb]
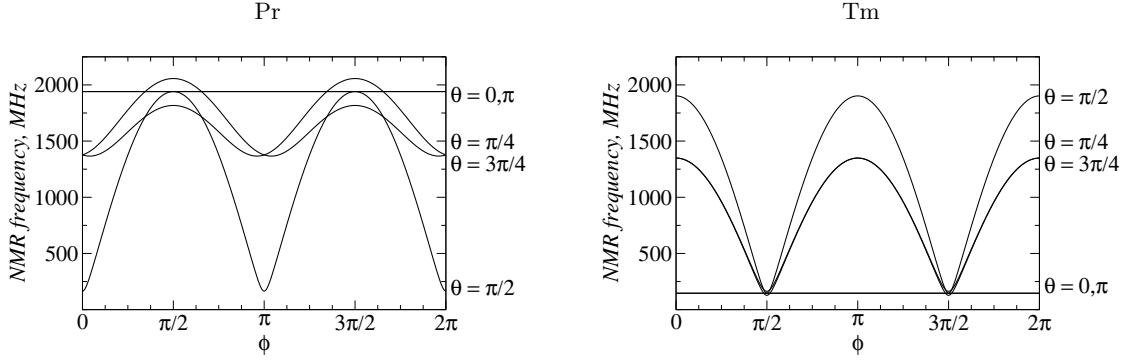

\begin{tabular}{c @{\hspace{1cm}} c}
Pr & Tm \\
&\\
\epsfig{figure=nmrPr_slc.eps,height=4cm} & \epsfig{figure=nmrTm_slc.eps,height=4cm}
\end{tabular}
\caption{
The NMR frequency versus the orientation of magnetic field with respect to the 
crystallographic axes, $B=10\ \textrm{T}$. We show the dependence on $\phi$ for different
values of $\theta$.}
\label{NMR}
\end{figure}

Equation (\ref{Hnuc}), together with (\ref{Jav}), gives the NMR frequency $\nu$.
Dependence of the frequency on the orientation of the magnetic field  
$\boldsymbol{B}=B(\sin\theta\cos\phi,\sin\theta\sin\phi,\cos\theta)$
with respect to the crystallographic axes is plotted in Fig.~\ref{NMR}, we take $B=10T$.
Equation (\ref{Hnuc}) also defines the quantization axis for the nuclear spin:
\begin{equation}
\boldsymbol{n}_I \propto \left( A_{\textrm{\it hf}} \langle \boldsymbol{J}\rangle
- \frac{\mu \mu_N}{I} \boldsymbol{B} \right)=
\left( A_{\textrm{\it hf}} \hat{\tau} \boldsymbol{B}
- \frac{\mu \mu_N}{I} \boldsymbol{B} \right).
\label{Iquant}
\end{equation}
This allows us to find cross product $\boldsymbol{n}_I\times\boldsymbol{J}$
that appears in the anapole induced energy correction (\ref{HdynI}), (\ref{de}):
\begin{equation}
M = |\boldsymbol{n}_I\times\boldsymbol{J}| = 
\frac{|\mu \mu_N[\boldsymbol{B}\times(\hat{\tau} \boldsymbol{B})]|}
{|I A_{\textrm{\it hf}} (\hat{\tau}\boldsymbol{B}) -\mu \mu_N \boldsymbol{B}|}.
\label{Sine}
\end{equation}
The value of $M$ depends on the magnitude and the orientation of the external magnetic field $\boldsymbol{B}$ with respect 
to the crystallographic axes. At $B=10\ \textrm{T}$ the maximum value of $M$ is
\begin{eqnarray}
\textrm{Pr}&:& M = 1.02 \cdot 10^{-1},\nonumber\\ 
\textrm{Tm}&:& M = 0.79 \cdot 10^{-1}.
\label{mdyntm}
\end{eqnarray}
Unfortunately, the values of $M$ are relatively small compared to the maximum possible value $M=J$ (4 for Pr and 6 for Tm).
The suppression is due to the fact that in the nuclear magnetic Hamiltonian (\ref{Hnuc}) the hyperfine interaction
$A_{\textrm{\it hf}} (\boldsymbol{J\cdot I})$ is an order of magnitude larger than the direct magnetic interaction
$\mu \mu_N (\boldsymbol{B\cdot I})/I$, while to maximize $M$ one has to have these interactions comparable.
In spite of the suppression, the observable effects related to the effective Hamiltonian (\ref{HdynI}), (\ref{de})
are quite reasonable (see next Section).

\begin{figure}[htb]
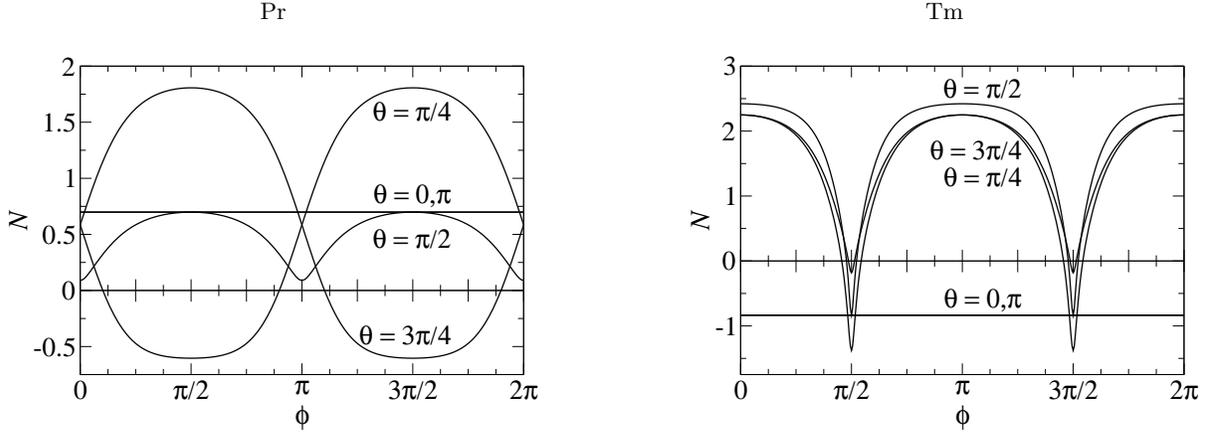

\begin{tabular}{c @{\hspace{2cm}} c}
Pr & Tm \\
&\\
\epsfig{figure=stPr_slc.eps,height=5cm} &
\epsfig{figure=stTm_slc.eps,height=5cm}
\end{tabular}
\caption{The kinematic coefficient $N$ (\ref{NNN}) in the lattice octupole induced energy 
correction versus orientation of magnetic field with respect to the crystallographic axes, $B=10\ \textrm{T}$.
We show the dependence on $\phi$ for different values of $\theta$.}
\label{JJJ}
\end{figure}

The situation with the effective interaction (\ref{HstI}), (\ref{effHam}) is different.
Looking at equations (\ref{HstI}), (\ref{effHam}) one can expect at first sight that the 
corresponding energy shift is nonzero only if $\boldsymbol{I}\times\langle \boldsymbol{J}\rangle \ne 0$. 
However, this is incorrect.
The point is that due to the crystal field the tensor $\langle J_j J_l J_m + J_m J_l J_j \rangle$
has nonzero components orthogonal to $\langle \boldsymbol{J}\rangle$. And the octupole induced 
energy shift is in fact maximum when $\boldsymbol{I} \ \Vert \ \langle \boldsymbol{J}\rangle$.
The dependence of the  kinematic coefficient (see Eq.~(\ref{effHam}))
\begin{equation}
\label{NNN}
N=\frac{1}{I}I_i \epsilon_{ijk} T_{klm} \langle J_j J_l J_m + J_m J_l J_j \rangle
\end{equation}
on the orientation of magnetic field  $\boldsymbol{B}=B(\sin\theta\cos\phi,\sin\theta\sin\phi,\cos\theta)$
at $B=10\ \textrm{T}$ is plotted in Fig.~\ref{JJJ}.  The maximum value of $N$ is
\begin{eqnarray}
\label{msttm}
\textrm{Pr}&:& \ \ \ N = 1.81,\nonumber\\ 
\textrm{Tm}&:& \ \ \ N = 2.42 .
\end{eqnarray}

The calculations in the present section are based on the fit of experimental energy levels, Table \ref{crfieldlev},
using the crystal field parameters. We use the set of parameters presented in Table \ref{crfieldpar}.
Unfortunately, the set is not unique and there are other sets which also reasonably fit the energy levels.
In particular, for Tm$^{3+}$ in YAG there is a set of parameters which gives a lattice octupole induced PNC energy 
shift an order of magnitude larger than the present set. At this stage we prefer to continue with the
conservative estimate. To elucidate the uncertainty related to the crystal field parameters 
detailed measurements of NMR frequencies, as well as transition amplitudes, are necessary.

\section{Estimates of observable effects}

The effect (\ref{HdynI}), (\ref{de}) requires a displacement of the impurity ion from its equilibrium position. 
Such displacement can be achieved by application of an external electric field.
The displacement has been estimated in Ref.~\cite{MDS} in relation to the discussion of electric dipole moments.
The idea behind the estimate is very simple.
Since the Ga--O link in YGG and the Al--O link in YAG are much more rigid than the Y--O links (see discussion in \cite{MDS})
the  electrostatic polarization in YGG and YAG is mainly due to displacement of the yttrium ions
\begin{equation}
\boldsymbol{P} = 3 e n \Delta\boldsymbol{r}.
\end{equation}
On the other hand, the dielectric polarization caused by the external electric field $\boldsymbol{E}$ is
\begin{equation}
\boldsymbol{P} = \frac{\epsilon - 1}{4\pi} \boldsymbol{E},
\end{equation}
where the static dielectric constant is $\epsilon \approx 12$ for YGG and YAG. This yields the following 
expression for the displacement of the yttrium ions:
\begin{eqnarray}
\label{dise}
\Delta\boldsymbol{r} &=& \frac{\epsilon - 1}{4\pi} \frac{\boldsymbol{E}}{3en},\nonumber \\
\Delta{r}/a_B &=& 3.0\cdot 10^{-8} E[\frac{\textrm{V}}{\textrm{cm}}] .
\end{eqnarray}
Measurements of infrared spectra, as well as
measurements of the dependence of the dielectric constant on the concentration of impurities, can help to improve
the estimate (\ref{dise}).

Using (\ref{de}), together with (\ref{mdyntm}) and (\ref{dise}), we obtain the
following estimates for the NMR frequency shift ($\Delta I = 1$) due to the nuclear anapole moment:
\begin{eqnarray}
\label{nuse}
\textrm{Pr}&:& \Delta\nu \sim 0.9 \times 10^{-9} E[\frac{\textrm{V}}{\textrm{cm}}]\ \textrm{Hz} , \nonumber\\
\textrm{Tm}&:& \Delta\nu \sim 0.5 \times 10^{-9} E[\frac{\textrm{V}}{\textrm{cm}}]\ \textrm{Hz} .
\end{eqnarray}

An alternative possibility for the experiment is to provide the maximum possible value of the cross  
product $\boldsymbol{I} \times \boldsymbol{J}$ by applying an RF pulse
and then to measure the induced electric field. Using (\ref{de}), together with estimates of  the elastic
constant with respect to the shift of the rare earth ion performed in \cite{MDS}, we arrive at the
following values of the anapole induced electric field:
\begin{eqnarray}
\label{ea}
\textrm{Pr}&:& E \sim 1.4 \times 10^{-6} \ \textrm{V/cm} , \nonumber\\
\textrm{Tm}&:& E \sim 0.4 \times 10^{-6} \ \textrm{V/cm} .
\end{eqnarray}
The field precesses around the direction of the magnetic field with a frequency of about 1 GHz
due to the nuclear spin precession. In the estimates (\ref{ea}) we assume that all yttrium ions
are substituted by the rare earth ions.

Another manifestation of nuclear anapole moment is the static perpendicular macroscopic magnetization induced 
by an external electric field, 
\begin{equation}
\delta \boldsymbol{I} \propto  \boldsymbol{B} \times \boldsymbol{E} .
\end{equation}
The exact value of the macroscopic magnetization depends on temperature and other experimental 
conditions, therefore we cannot present a specific value. However, we can compare the effect
with that expected in the electron EDM experiment \cite{Lam} (correlation 
$\delta \boldsymbol{J} \propto \boldsymbol{E}$) using  the present experimental limit 
on $d_e$ \cite{Com}, $d_e = 1.6\times 10^{-27} e \ \textrm{cm}$, as a reference point.
The effective anapole interaction (\ref{de}) is four order of magnitude larger
than the similar effective EDM interaction \cite{MDS}. On the other hand, the 
electron EDM interaction causes electron magnetization whereas the anapole interaction
causes only nuclear magnetization, so we lose 3 orders of magnitude on the value of the
magnetic moment. Therefore, altogether, one should expect that the anapole magnetization is several
times larger than the EDM magnetization.

The effective interaction (\ref{effHam}) is independent of the external electric field and is due to 
the asymmetric environment of the rare earth ion site. Since there is always another site within the unit cell which is
the exact mirror reflection of the first one, the energy correction (\ref{effHam}) does actually lead to the NMR line splitting.
Using Eqs.~(\ref{effHam}), (\ref{eq45}), (\ref{NNN}), and (\ref{msttm}), we find the maximum value of this splitting corresponding to the magnetic field $B = 10\ \textrm{T}$:
\begin{eqnarray}
\textrm{Pr}&:& \Delta\nu \sim 0.5 \ \textrm{Hz} , \nonumber\\
\textrm{Tm}&:& \Delta\nu \sim 0.25\ \textrm{Hz} .
\end{eqnarray}
The splitting depends on the orientation of the magnetic field with respect to
the crystallographic axes, see Fig.~\ref{NMR}.

\section{Conclusions}
In the present work we have considered effects caused by the nuclear anapole moment in
thulium doped yttrium aluminium garnet and praseodymium doped yttrium gallium garnet. There are two 
effects related to the frequency of NMR: 1) NMR line shift in combined electric and magnetic fields. 
The shift is about $10^{-5}$ Hz at $B=10$ T and $E=10$ kV/cm. 2) NMR line splitting 
(magnetic field only). The spitting is about 0.5 Hz at $B=10$ T.
The value of the splitting depends on the orientation of the magnetic field with respect to
the crystallographic axes.
Another PNC effect is the induced RF electric field orthogonal to the plane of the magnetic field 
and nuclear spin, $\boldsymbol{E} \propto [\boldsymbol{B} \times \boldsymbol{I}]$.
The field is $E \sim 10^{-6}\ \textrm{V/cm}$ at magnetic field $B=5$--10 T.
The last effect we have discussed is unrelated to NMR. This is a variation of the static
macroscopic magnetization in combined electric and magnetic fields, 
$\delta \boldsymbol{M} \propto \boldsymbol{B} \times \boldsymbol{E}$.
The magnitude of the effect is several times larger than that expected in the electric dipole moment
experiment \cite{Lam}.

It is our pleasure to acknowledge very helpful discussions with D. Budker, V.V. Yashchuk, A.O. Sushkov and A.I. Milstein.

\section{Appendix. Radial equations}

In order to calculate the energies and wavefunctions of unperturbed states of the single impurity ion in 
the garnet environment, we use the Dirac equation
\begin{equation}
\label{Deq}
(H-\epsilon)|\psi\rangle = 0.
\end{equation}
The effective potential $V(r)$ (\ref{pot2}) in the Dirac Hamiltonian $H$ is spherically 
symmetric, and thus the two-component wavefunction $|\psi\rangle$ is of the form
\begin{equation}
\label{psi}
|\psi\rangle=\frac{1}{r}{f(r)\Omega_{\kappa} \choose i\alpha g(r) \widetilde{\Omega}_{\kappa}}.
\end{equation}
Here $\Omega_{\kappa}$ and $\widetilde{\Omega}_{\kappa}$ are the spherical spinors and $f(r)$ and $g(r)$ are 
radial wavefunctions.
Substituting expression (\ref{psi}) for $|\psi\rangle$ into the Dirac equation (\ref{Deq}), 
one gets the following radial equations
\begin{eqnarray}
&&f^{\prime} + \kappa f/x + (-2 + \alpha^2(V-\epsilon))g=0,\nonumber\\
&&g^{\prime} - \kappa g/x - (V-\epsilon)f=0.
\label{fgeqs}
\end{eqnarray}
Here $x=r/a_B$ is the radius in atomic units; $\kappa=(-1)^{j+1/2-l}(j+1/2)$, where $j$ and $l$ are 
the total and orbital angular momenta of the single-electron state correspondingly; the potential $V(x)$ 
,as well as the energy $\epsilon$, is expressed in atomic energy units.
Solving the system of equations (\ref{fgeqs}) as an eigenvalue problem numerically on a logarithmic 
coordinate grid, we find energies and wavefunctions of the unperturbed states.

The inhomogeneous Dirac equations (\ref{dp2}) and (\ref{dp3}) are of the form
\begin{equation}
\label{Seq}
(H-E)|\delta\psi\rangle = -\hat{V}_p|\psi\rangle,
\end{equation}
where $\hat{V}_p$ is the single-particle perturbation operator.
The correction $|\delta \psi\rangle$ is of the form
\begin{equation}
\label{dpsi}
|\psi\rangle=\frac{1}{r}{F(r)\Omega_{\kappa'} \choose i\alpha G(r) \widetilde{\Omega}_{\kappa'}},
\end{equation}
and hence the corresponding radial equations are
\begin{eqnarray}
 &&F^{\prime} + \kappa' F/x + (-2 + \alpha^2(V-\epsilon))G=R_f 
\langle\Omega_{\kappa'}|\hat{\Phi}|\Omega_{\kappa}\rangle,\nonumber\\
&&G^{\prime} - \kappa' G/x - (V-\epsilon)F=R_g \langle
\tilde{\Omega}_{\kappa'}|\hat{\Phi}|\tilde{\Omega}_{\kappa}\rangle .
\label{corr}
\end{eqnarray}
The operator $\hat{\Phi}$ represents the angular part of the perturbation $\hat{V}_p$,
and $R_f$ and $R_g$ are the radial parts of the perturbation.
The functions $R_f$, $R_g$, and $\hat{\Phi}$ for all the cases we need in the present work
are presented in Table \ref{tab2}. 
\begin{table}
\begin{tabular}{c@{\hspace{0.5cm}}c@{\hspace{0.5cm}}c@{\hspace{0.5cm}}c@{\hspace{0.5cm}}c}
\hline
&$\hat{V}_p=V_a,(\ref{pert1})$
&$\hat{V}_p=V_a,(\ref{pert1})$
&$\hat{V}_p=V_1,(\ref{pert2})$
&$\hat{V}_p=V_{3},(\ref{pert3m})$
\\
&$|\psi\rangle=|ns_{1/2}\rangle$
&$|\psi\rangle=|np_{1/2}\rangle$
&$|\psi\rangle=|ns_{1/2}\rangle$ or $|\psi\rangle=|np_{1/2}\rangle$
&$|\psi\rangle=|ns_{1/2}\rangle$ or $|\psi\rangle=|np_{1/2}\rangle$
\\ 
\hline $R_f$ 
&$-K_a \alpha \rho_n(x) f(x)$            
&$-\frac{1}{3} K_a \alpha \rho_n(x)f(x)$ 
&$-2\frac{(r-r_o)}{D^2}A_o e^{-\left(\frac{r-r_o}{D}\right)^2}\alpha^2g(x)$
&$A_o e^{-\left(\frac{r-r_o}{D}\right)^2}\alpha^2g(x)$
\\ 
$R_g$ 
&$\frac{1}{3} K_a \alpha \rho_n(x)g(x)$  
&$K_a \alpha \rho_n(x)g(x)$
&$2\frac{(r-r_o)}{D^2}A_o e^{-\left(\frac{r-r_o}{D}\right)^2} f(x)$
&$-A_o e^{-\left(\frac{r-r_o}{D}\right)^2} f(x)$
\\ 
\hline $\hat{\Phi}$
&$-2i(\boldsymbol{I \cdot j})$
&$2i(\boldsymbol{I \cdot j})$
&$\Delta x \sin\theta \cos\phi +
\Delta y \sin\theta\sin\phi
+ \Delta z \cos\theta$
&$\frac{\pi}{2} T_{3m}\cdot Y_{3m}(\boldsymbol{r})$
\\ 
\hline 
\end{tabular}
\caption{The functions $R_f$, $R_g$ and $\hat{\Phi}$ for the different perturbation operators 
and different  states $|\psi\rangle$. $\rho_n$ is nuclear density normalized to unity.}
\label{tab2}
\end{table}

\begin{table}
\begin{tabular}{l c @{\hspace{0.5cm}} c}
\hline
Diagram & Pr$^{3+}$ & Tm$^{3+}$ \\
\hline \vspace{-7pt}\\
\multicolumn{3}{c}{Dipole effect}\\
\vspace{-7pt}\\
\vspace{6pt} 1,7,8,11
&$\frac{2^2 \cdot 43}{3^2 \cdot 5^3 \cdot 7}F(2)
 -\frac{2 \cdot 19}{3^6 \cdot 5 \cdot 7}F(4)$
&$\frac{1}{2 \cdot 3^2 \cdot 5 \cdot 7}F(2)
 -\frac{79}{2 \cdot 3^6 \cdot 7}F(4) $\\
\vspace{6pt} 2,9,10
&$\frac{2^2 \cdot 23}{3^2 \cdot 5^3 \cdot 7}F(2)
 +\frac{2 \cdot 5}{3^6 \cdot 7}F(4)
$
&$-\frac{1}{2 \cdot 3^2 \cdot 5 \cdot 7}F(2)
  -\frac{2 \cdot 17}{3^6 \cdot 7}F(4) $\\
\vspace{6pt} 3,4,5,6
&$\frac{2}{3 \cdot 5 \cdot 7}F(3) $
&$-\frac{1}{3^2 \cdot 7}F(3) $\\
\vspace{-7pt}\\
\multicolumn{3}{c}{Lattice octupole effect}\\
\vspace{-7pt}\\
\vspace{6pt} 1,2,3
&$ T_{31}\sqrt{\frac{\pi}{21}} \frac{13 \cdot F(3)}{2 \cdot 3^2 \cdot 5^2 \cdot 7 \cdot 11}$
&$-T_{31}\sqrt{\frac{\pi}{21}} \frac{F(3)}{2^2 \cdot 3 \cdot 5 \cdot 7 \cdot 11}$
\\
\vspace{6pt} 4,5
&$-T_{31}\sqrt{\frac{\pi}{21}} \left[
 \frac{13 \cdot 29 \cdot F(2)}{3^2 \cdot 5^3 \cdot 7^2 \cdot 11}
+\frac{5 \cdot 13 \cdot F(4)}{2 \cdot 3^4 \cdot 7^2 \cdot 11^2}\right]$
&$ T_{31}\sqrt{\frac{\pi}{21}} \left[
 \frac{F(2)}{2 \cdot 3 \cdot 5^2 \cdot 7 \cdot 11}
-\frac{F(4)}{2^2 \cdot 3^3 \cdot 5 \cdot 7 \cdot 11^2}\right]$
\\
\vspace{6pt} 6,7
&$ T_{31}\sqrt{\frac{\pi}{21}} \left[
 \frac{13^2 \cdot F(2)}{2^2 \cdot 3 \cdot 5^3 \cdot 7^2 \cdot 11}
+\frac{13 \cdot 47 \cdot F(4)}{2^2 \cdot 3^3 \cdot 5 \cdot 7^2 \cdot 11^2}\right]$
&$-T_{31}\sqrt{\frac{\pi}{21}} \left[
 \frac{F(2)}{2 \cdot 3 \cdot 5^2 \cdot 7 \cdot 11}
+\frac{2^2 \cdot F(4)}{3^2 \cdot 5 \cdot 7 \cdot 11^2}\right]$
\\
\hline
\end{tabular}
\caption{
{\it Dipole effect:} Angular coefficient for each of the 11 diagrams shown in Fig.~\ref{alld}. 
The factor $(\Delta x I_y - \Delta y I_x)J_z$, which corresponds to the kinematic structure (\ref{de}) 
and which is common for all the contributions, is omitted.
$F(l)$ denotes the Coulomb integral of multipolarity $l$ in the radial part of the diagam. 
\newline
{\it Lattice octupole effect:} Angular coefficients for each of the 7 diagrams shown in Fig.~\ref{Y3diags}. 
The factor $IJ_z[5J_z^2-3J(J+1)+1]$, which corresponds to the kinematic structure (\ref{effHam}) 
and which is common for all the contributions, is omitted.
}
\label{ang}
\end{table}

Having separated the radial parts, one can calculate the angular coefficients for the diagrams in 
Figs.~\ref{alld} and \ref{Y3diags}. The results of these calculations are presented in Table \ref{ang}.
The electronic configurations of Pr$^{3+}$ and Tm$^{3+}$ are similar:
two $f$-electrons in Pr$^{3+}$ and two $f$-holes in Tm$^{3+}$. However, their orbital and spin angular momenta combine to yield different
total angular momenta, and this makes the angular coefficients for Pr$^{3+}$ and Tm$^{3+}$ different.

\end{document}